\documentclass[twoside,twocolumn,9pt]{article}
\usepackage{extsizes}
\usepackage[super,sort&compress,comma]{natbib} 
\usepackage[version=3]{mhchem}
\usepackage[left=1.5cm, right=1.5cm, top=1.785cm, bottom=2.0cm]{geometry}
\usepackage{balance}
\usepackage{times,mathptmx}
\usepackage{sectsty}
\usepackage{graphicx} 
\usepackage{lastpage}
\usepackage[format=plain,justification=justified,singlelinecheck=false,font={stretch=1.125,small,sf},labelfont=bf,labelsep=space]{caption}
\usepackage{float}
\usepackage{fancyhdr}
\usepackage{fnpos}
\usepackage[english]{babel}
\addto{\captionsenglish}{%
  \renewcommand{\refname}{Notes and references}
}
\usepackage{array}
\usepackage{droidsans}
\usepackage{charter}
\usepackage[T1]{fontenc}
\usepackage[usenames,dvipsnames]{xcolor}
\usepackage{setspace}
\usepackage[compact]{titlesec}
\usepackage[hidelinks]{hyperref}
\usepackage{amsmath,amssymb}
\usepackage{multirow}
\usepackage{wasysym}

\usepackage{epstopdf}

\definecolor{cream}{RGB}{222,217,201}

\begin{document}

\pagestyle{fancy}
\thispagestyle{plain}
\fancypagestyle{plain}{

\renewcommand{\headrulewidth}{0pt}
}

\makeFNbottom
\makeatletter
\renewcommand\LARGE{\@setfontsize\LARGE{15pt}{17}}
\renewcommand\Large{\@setfontsize\Large{12pt}{14}}
\renewcommand\large{\@setfontsize\large{10pt}{12}}
\renewcommand\footnotesize{\@setfontsize\footnotesize{7pt}{10}}
\makeatother

\renewcommand{\thefootnote}{\fnsymbol{footnote}}
\renewcommand\footnoterule{\vspace*{1pt}%
\color{cream}\hrule width 3.5in height 0.4pt \color{black}\vspace*{5pt}} 
\setcounter{secnumdepth}{5}

\makeatletter 
\renewcommand\@biblabel[1]{#1}            
\renewcommand\@makefntext[1]%
{\noindent\makebox[0pt][r]{\@thefnmark\,}#1}
\makeatother 
\renewcommand{\figurename}{\small{Fig.}~}
\sectionfont{\sffamily\Large}
\subsectionfont{\normalsize}
\subsubsectionfont{\bf}
\setstretch{1.125} 
\setlength{\skip\footins}{0.8cm}
\setlength{\footnotesep}{0.25cm}
\setlength{\jot}{10pt}
\titlespacing*{\section}{0pt}{4pt}{4pt}
\titlespacing*{\subsection}{0pt}{15pt}{1pt}

\newcommand{\kB}{\mathrm{k}_\mathrm{B}}
\newcommand{\kQTST}{k_\mathrm{QTST}}
\newcommand{\kRPMD}{k_\mathrm{RPMD}}
\newcommand{\kTIQM}{k_\mathrm{TIQM}}
\newcommand{\kSQM}{k_\mathrm{SQM}}
\newcommand{\kST}{k_\mathrm{ST}}
\newcommand{\kQCT}{k_\mathrm{QCT}}
\newcommand{\kappap}{\kappa_\mathrm{plateau}}

\fancyfoot{}
\fancyhead{}
\renewcommand{\headrulewidth}{0pt} 
\renewcommand{\footrulewidth}{0pt}
\setlength{\arrayrulewidth}{1pt}
\setlength{\columnsep}{6.5mm}
\setlength\bibsep{1pt}

\makeatletter 
\newlength{\figrulesep} 
\setlength{\figrulesep}{0.5\textfloatsep} 

\newcommand{\topfigrule}{\vspace*{-1pt}%
\noindent{\color{cream}\rule[-\figrulesep]{\columnwidth}{1.5pt}} }

\newcommand{\botfigrule}{\vspace*{-2pt}%
\noindent{\color{cream}\rule[\figrulesep]{\columnwidth}{1.5pt}} }

\newcommand{\dblfigrule}{\vspace*{-1pt}%
\noindent{\color{cream}\rule[-\figrulesep]{\textwidth}{1.5pt}} }

\makeatother

\twocolumn[
  \begin{@twocolumnfalse}
\sffamily
\begin{tabular}{m{0.0cm} p{18.0cm} }

& \noindent\LARGE{\textbf{The low temperature D$^+$ + H$_2 \rightarrow$ HD + H$^+$ reaction rate coefficient: a ring polymer molecular dynamics and quasi-classical trajectory study}} \\
\vspace{0.3cm} & \vspace{0.3cm} \\

& \noindent\large{Somnath~Bhowmick,$^{\ast}$\textit{$^{a}$} Duncan~Bossion,\textit{$^{b}$} Yohann~Scribano,\textit{$^{b}$} and Yury~V.~Suleimanov\textit{$^{a}$}} \\

\vspace{0.3cm} & \vspace{0.3cm} \\
& \noindent\normalsize{The reaction between D$^+$ and H$_2$ plays an important role in astrochemistry at low temperatures and also serves as a prototype for simple ion-molecule reaction. Its ground $\tilde{X}~^1A^{\prime}$ state  has a very small thermodynamic barrier (up to 1.8$\times 10^{-2}$ eV) and the reaction proceeds through the formation of an intermediate complex lying within the potential well of depth of at least 0.2 eV thus representing a challenge for dynamical studies.  In the present work, we analyze the title reaction within the temperature range of 20 $-$ 100 K by means of ring polymer molecular dynamics (RPMD) and quasi-classical trajectory (QCT) methods over the full-dimensional global potential energy surface developed by Aguado {\textit{et al.}} [A. Aguado, O. Roncero, C. Tablero, C. Sanz, and M. Paniagua, {\textit{J. Chem. Phys.}}, 2000, {\bf{112}}, 1240]. The computed thermal RPMD and QCT rate coefficients are found to be almost independent of temperature and fall within the range of 1.34 $-$ 2.01$\times$10$^{-9}$ cm$^3$ s$^{-1}$. They are also  in a very good agreement with the previous time-independent quantum mechanical and statistical quantum method calculations. Furthermore, we observe that  the choice of asymptotic  separation distance between the reactants can markedly alter the rate coefficient in the low temperature regime (20 $-$ 50 K). Therefore it is of utmost importance to correctly assign the value of this parameter for dynamical studies, particularly at very low temperatures of astrochemical importance. We finally conclude that experimental rate measurements for the title reaction are highly desirable in future.} \\

\end{tabular}

 \end{@twocolumnfalse} \vspace{0.6cm}

  ]

\renewcommand*\rmdefault{bch}\normalfont\upshape
\rmfamily
\section*{}
\vspace{-1cm}


\footnotetext{\textit{$^{a}$~Computation-based Science and Technology Research Center, The Cyprus Institute, 20 Konstantinou Kavafi Street, Nicosia 2121, Cyprus. Fax: +357 22208625; Tel: +357 22397526; E-mail: s.bhowmick@cyi.ac.cy}} 
\footnotetext{\textit{$^{b}$~Laboratoire Univers et Particules de Montpellier, UMR-CNRS 5299, Universit\'{e} de Montpellier, Place Eug\`ene Bataillon, 34095 Montpellier, France.}}


\section{Introduction}\label{sec:introduction}
Primordial gases such as the deuterated hydrogen molecule, HD, plays an important role in the field of astrochemistry particularly during the early universe epoch. For example, the cooling of HD molecule is linked to the gravitational collapse and the fragmentation of clouds.\cite{Dalgarno_AJL_1972, Flower_MNRAS_2001, Dalgarno_JPBAMOP_2002,McGreer_AJ_2008} The role of HD in radiative cooling of post-shock gas has been mentioned in the work by Flower and his co-worker.\cite{Flower_MNRAS_2001} Shocks must occur during the collapse of gas in the early universe and is responsible for biased galaxy and star formation.\cite{Dalgarno_JPBAMOP_2002} The cooling of HD can be more pronounced than abundant H$_2$ molecule specifically at low temperatures, since it possesses a small dipole moment ($\approx$ 8.3$\times$10$^{-4}$ D\cite{Abgrall_AASS_1982}) and therefore the low energy transition $\Delta J=\pm 1$ is allowed, leading to a greater transition rate.\cite{Dalgarno_JPBAMOP_2002, Flower_MNRAS_2001} Quantum calculation of Flower {\textit{et al.}}\cite{Flower_MNRAS_2000} showed that the cooling function (rate of cooling) of HD is much larger than that for H$_2$. HD can allow gases to cool within $\sim$300 K to $\lesssim$ 100 K and is important in star formation in low mass halos (few times of 10$^5$ M$_{\odot}$).\cite{McGreer_AJ_2008} Despite the fact that the cosmic abundance of D is very low (D/H $\approx$ 10$^{-5}$), the fractional abundance of HD can be enhanced to about 2 orders of magnitude compared to H$_2$ by chemical fractionation.\cite{Puy_AA_1993, Galli_AA_1998, Glover_MNRAS_2008, Flower_MNRAS_2000_2} Uehara and his co-worker\cite{Uehara_AJ_2000} found that almost all deuterium can be converted to HD molecule and subsequent cooling may lead to the formation of primordial low mass stars and brown dwarfs. 

In the diffuse interstellar medium, HD can be obtained by deuteron exchange with H$_2$:\cite{Dalgarno_AL_1973, Watson_AJ_1973, Lipshtat_MNRAS_2004, Glover_MNRAS_2008, Snow_AJ_2008}
\begin{equation}\label{eq:rxn_HD}
\text{D}^++\text{H}_2 \to \text{HD}+\text{H}^+.
\end{equation}
At energies below 1.8 eV, the proton-deuteron exchange is the only open reactive channel.\cite{Gerlich_ACP_1992} In addition, there are several other ways to generate HD in the atmosphere, such as the radiative association reaction (H + D $\to$ HD + $\nu$),\cite{Stancil_AJ_1997,Dalgarno_JPBAMOP_2002} dissociative recombination of H$_2$D$^+$ (H$_2$D$^+$ + $e^-$ $\to$ HD + H)\cite{Dalgarno_AA_1993} and at  redshift $z\lesssim10$ via charge exchange (HD$^+$ + H $\to$ HD + H$^+$).\cite{Karpas_JCP_1979} Cosmological model of Dalgarno {\textit{et al.}}\cite{Dalgarno_AL_1973} have shown that the reaction between H$_2$ and D$^+$ is the primary source of HD and the others may only give rise to minor contribution. The reaction (1) is exoergic by 39.8 meV and has no potential barrier,\cite{Henchman_JCP_1981} while HD destruction by the reverse reaction (HD + H$^+$ $\to$ D$^+$ + H$_2$) is endothermic by 82.9 meV\cite{Glover_MNRAS_2008} and therefore HD/H$_2$ ratio can build up-to two-fold at low temperature regime ($<$ 200 K). The most efficient condition for the formation of HD is in the gas phase with the temperature below 150 K. Above this temperature, the production of HD can be enhanced on grain surfaces, particularly in metal-poor environments (with metallicities as low as 10$^{-5}$ Z$_{\odot}$).\cite{Cazaux_AA_2009}

The title reaction also serves as a prototype for simplest ion-molecule reaction\cite{Scribano_JPCA_2013} and therefore attracted considerable interest to study its dynamical properties, both experimentally\cite{Fehsenfeld_JCP_1974, Henchman_JCP_1981, Henchman_JCP_1982, Alge_APJ_1982, Gerlich_Sympo_1982, Gerlich_ACP_1992} and theoretically.\cite{Markovic_CP_1995, Takayanagi_JCP_2000, Kamisaka_JCP_2002, Gerlich_PSS_2002, Lezana_JCP_2005, Varandas_CPL_2009, Jambrina_PCCP_2010, McCarroll_PS_2011, Jambrina_PCCP_2012, Scribano_JPCA_2013, Scribano_JCP_2013, Scribano_JPCA_2014, Jambrina_PRA_2015, Jambrina_JCP_2015, Varandas_JCP_2015} As a consequence, the dynamics of the low energy H$-$D exchange processes is well understood. In the following, we summarise the previous reports on the dynamical studies of the D$^+$ + H$_2$ reaction.

One of the earliest experimental investigations on the rate coefficient of the title reaction was due to Henchman and his co-workers by temperature variable selected ion flow tube (SIFT)\cite{Henchman_JCP_1981} and drift tube\cite{Henchman_JCP_1982} methods. For the SIFT study, the measured rate coefficients at selected temperatures at 205 K (2.2$\pm$0.1$\times$10$^{-9}$ cm$^3$ s$^{-1}$) and 295 K (1.7$\pm$0.1$\times$10$^{-9}$ cm$^3$ s$^{-1}$) are close to the Langevin value (2.1$\times$10$^{-9}$ cm$^3$ s$^{-1}$) and is corroborated by the SIFT investigation of  Smith {\textit{et al.}}\cite{Alge_APJ_1982} The authors also pointed out the role of statistical factors in kinetics in the sense that all collision of reaction (\ref{eq:rxn_HD}) can lead to observable product. The drift tube analysis at 295 K yields the rate coefficient, 1.1$\times$10$^{-9}$ cm$^3$ s$^{-1}$, that differs considerably than the one obtained by SIFT. Prior to the work of Henchman and his co-workers, Fehsenfeld {\textit{et al.}} also determined the rate coefficient within 80 $-$ 278 K by flowing afterglow (FA) apparatus.\cite{Fehsenfeld_JCP_1974} In 1982\cite{Gerlich_Sympo_1982} (resp. 2002\cite{Gerlich_PSS_2002}), Garlich used the most dynamically biased (MDB) statistical theory (ST) based on some analytical function to calculate the rate coefficients within the temperature range of 30 $-$ 600 K (resp. 30 $-$ 130 K). At 295 K, the calculated rate coefficient, 1.69$\times$10$^{-9}$ cm$^3$ s$^{-1}$, is in good agreement with the experimental SIFT results of Henchman {\textit{et al.}}\cite{Henchman_JCP_1981} However, the authors suggested that this agreement possibly arises from an experimental error. Later on, Garlich used merged beams technique and found that the rate coefficient ($\sim$1.6$\times$10$^{-9}$ cm$^3$ s$^{-1}$) is independent of temperature (180 $-$ 350 K) with a typical Langevin cross-section.\cite{Gerlich_ACP_1992}

Theoretically, in 2005, Gonz\'{a}lez-Lezana {\textit{et al.}}\cite{Lezana_JCP_2005} studied the reactive non-charge transfer ion-molecule collisions D$^+$ + H$_2$ reaction using statistical quantum method (SQM) and different wave packet approaches. They found that the SQM method can reproduce the exact reaction probabilities for zero total angular momentum ($J$), but tend to differ for $J>0$. They also proposed different centrifugal sudden approximations for $J>0$ to overcome the shortcomings of the usual centrifugal sudden approach and calculated the integral and differential cross-sections. Later on, Jambrina {\textit{et al.}}\cite{Jambrina_PCCP_2010, Jambrina_PCCP_2012, Jambrina_PRA_2015, Jambrina_JCP_2015} reported on the theoretical dynamics of the title reaction. First,\cite{Jambrina_PCCP_2010} they determined the reaction probabilities for collisional energy range from 4 meV to 0.2 eV  by time-independent quantum mechanical (TIQM) calculation using the close-coupled hyperspherical method of Skouteris {\textit{et al.}}\cite{Skouteris_CPC_2000} on the ground $\tilde{X}~^1A^{\prime}$ potential energy surface (PES) of Aguado {\textit{et al.}}\cite{Aguado_JCP_2000} (developed using full configuration interaction method with extended basis set) for $J=0-40$. They show that the TIQM results agree relatively well with the approximate statistical quasi-classical trajectory (SQCT) model. In their follow-up paper,\cite{Jambrina_PCCP_2012} they compared the thermal rate coefficients (within 100 $-$ 500 K) calculated by TIQM, SQCT and QCT methods with the experimental measurements using either FA\cite{Fehsenfeld_JCP_1974} or SIFT\cite{Henchman_JCP_1981} techniques. Their calculation indicates that the rate coefficients do not depend significantly on the temperature. The SQCT rate coefficients agree very well with the experiment at 295 K, while TIQM result is about $\sim$15\% lower than the measured value. In addition, they have also calculated the state specific rate coefficients as a function of the translational energy ($E_T$) up-to 1.2 eV and found that for $j=0-3$, the rate coefficients of the three theoretical methods grow weakly with the increasing $E_T$. More recently, the rate coefficients were calculated within a very extensive temperature range, from deep ultracold (10$^{-8}$ K) regime to the Langevin (150 K) one, by either hyperspherical reactive scattering method\cite{Launay_CPL_1990} or modified version it.\cite{Jambrina_PRA_2015, Jambrina_JCP_2015} These calculations were based on the PES of Velilla {\textit{et al.}} (VLABP PES)\cite{Velilla_JCP_2008} which can accurately reproduce the long-range interactions and is a modification of the PES of Aguado {\textit{et al.}} (ARTSP PES).\cite{Aguado_JCP_2000} In these calculations, the reaction rate coefficients changed only within one order of magnitude while the collision energy changed within ten orders of magnitude. The rate coefficients are in very good agreement with the experiment.\cite{Gerlich_ACP_1992}

Recently, one of us (Y. Scribano) carried out time-independent state-to-state quantum mechanical study to calculate the rate coefficients at low temperature regime (up-to 100 K) using TIQM and SQM methods.\cite{Scribano_JPCA_2013, Scribano_JCP_2013, Scribano_JPCA_2014} These calculations are based on VLABP PES. The thermal rate coefficients are found to be almost independent of temperature and they are in a very good agreement with the experiment.\cite{Scribano_JCP_2013}  The state-to-state TIQM rate coefficient is very sensitive to translational energy $E_T$ due to large number of resonances. The correspondence between the TIQM and SQM state specific rate coefficients for D$^+$ + H$_2$ ($\nu$ = 0, $j$ = 0, 1) reaction found to be acceptable\cite{Scribano_JCP_2013} but the subsequent study reveals that the SQM rate coefficients slightly overestimates the TIQM.\cite{Scribano_JPCA_2014} The TIQM rate coefficients\cite{Scribano_JCP_2013} for reaction initiated at $j$ = 0 and $j$ = 1 is smaller and larger respectively than those obtained by the statistical model of McCarroll.\cite{McCarroll_PS_2011} McCarroll calculated the rate coefficient for the formation of HD within the temperature range 10 $-$ 400 K by statistical mixing model including nuclear symmetry constraint and assuming that the reaction proceeds via the formation of a long-lived complex that can decay into an energetically accessible reactive product with a high probability. McCarroll's calculations are in excellent agreement with the experiment\cite{Henchman_JCP_1981} and can reproduce both the magnitude and the temperature dependence of the reaction rate at temperature 205 K and 295 K.

To the best of our knowledge, the experimental rate coefficient of cosmologically important title reaction is still unknown at low temperature ($<$100 K). The uncertainty about the rate coefficients at low temperature may affect the predictions for HD abundance in the interstellar medium and consequently its cooling rate. One of the method that is able to calculate the rate coefficient of such bimolecular reactions and moreover exhibited excellent agreement with the experiment in previous studies of similar insertion reactions\cite{Suleimanov_JCP_2014, Suleimanov_JPCL_2014, Suleimanov_JPCL_2015, Suleimanov_JPCA_rev_2016} is the ring polymer molecular dynamics (RPMD) approximation.\cite{Manolopoulos_JCP_2004, Manolopoulos_Annu_Rev_Phys_Chem_2013}
The RPMD method has become very popular in recent years due to its accuracy and robustness.\cite{Suleimanov_JCP_2009, Suleimanov_JCP_2011, Suleimanov_JPCL_2012, Suleimanov_JCP_2013, Suleimanov_JCP_2013_1, Suleimanov_JPCL_2013, Suleimanov_PCCP_2013, Suleimanov_JCP_2014, Suleimanov_JPCL_2014, Suleimanov_JPCL_2014_1, Suleimanov_PCCP_2014, Suleimanov_JPCA_2014, Suleimanov_JPCA_2014_1, Suleimanov_JPCA_2014_2, Suleimanov_JCP_2015, Suleimanov_JPCB_2015, Suleimanov_JPCL_2015, Suleimanov_JPCA_rev_2016, Suleimanov_JPCB_2016, Suleimanov_JPCA_2016, Suleimanov_JPCA_2017, Suleimanov_PCCP_2017, Suleimanov_PCCP_2017_2, Suleimanov_PCCP_2017_3, Suleimanov_JPCL_2018,Suleimanov_SA_2018} In this present study, we determine the thermal rate coefficients of the D$^+$ + H$_2$ reaction by employing the RPMD rate theory at very low temperatures (20 $-$ 100 K) and compare with the previous experimental and theoretical results. 

The paper is organized as follows: in Section~\ref{sec:method} we provide the details of the QCT method and RPMD approach and its application to chemical reactions in the gas phase along with the PES used in the present study. In Section~\ref{sec:result} we discuss the results of RPMD and QCT rate coefficients and compare these with the earlier studies.\cite{Scribano_JPCA_2014,Gerlich_PSS_2002} Concluding remarks are given in Section~\ref{sec:conclusion}.

\section{Theoretical methods}\label{sec:method}
\subsection{Quasi-classical trajectory method}
Quasi-classical trajectory (QCT) calculations have been performed for the D$^+$ + H$_2(v=0,j=0)$ are based on the adiabatic full-dimensional global PES for the ground $\tilde{X}~^1A^{\prime}$ state of the H$_3^+$ system, developed by Aguado and co-workers (ARTSP PES).\cite{Aguado_JCP_2000} The analytical PES is constructed from 8469 {\textit{ab initio}} points calculated at the full configuration interaction (FCI) level of theory using 11$s$6$p$2$d$ (uncontracted)/8$s$6$p$2$d$ (contracted) basis functions. The DIM approach, corrected by many symmetrized three-body terms with 96 linear parameters and 3 nonlinear parameters, has been used to fit the global PES. The root-mean-square (RMS) error of the ARTSP PES found to be lower than 20 cm$^{-1}$. When some points of the ARTSP PES compared with the exact Born-Oppenheimer,\cite{Klopper_JCP_1994} they tend to mutually agree with each other. For example, the ARTSP PES energy for H$_2$ + H$^+$ dissociation is just 15 cm$^{-1}$ smaller than the exact energy. ARTSP PES has an equilateral triangle geometry at its minimum point, surrounded by a ($\approx$ 4 eV) deep well. Infrared spectra for transition restricted to this bound state has an accuracy that falls within few wave numbers.\cite{Lezana_JCP_2005} The entrance channel is devoid of a potential barrier, so the real-time dynamics is expected to be dominated by long-lived resonances. The reader is referred to ref.~\citenum{Aguado_JCP_2000} for further details on the ARTSP PES.

The QCT calculations are done following the same procedure described in ref.~\citenum{DuncanMNRAS}. Since this process is exothermic, low collisional energies are required to converge the reaction cross-section and chemical rate coefficient. The threshold collisional energy is at $9.95\times10^{-6}$~Hartree, corresponding to $2.45\times10^{-4}$~eV (or $3.14$~K).  We select a grid of 37 energy points distributed in a judicious way to correctly cover the low collisional energy range. We selected energy points every $2.38\times10^{-4}$~Hartree ($6.48\times10^{-3}$~eV) until $5.01\times10^{-3}$~Hartree ($1.36\times10^{-2}$~eV). As the cross section is more monotonic above, we take 15 energy points every $5.18\times10^{-3}$~Hartree ($1.41\times10^{-1}$~eV) up to $2.96\times10^{-2}$~Hartree ($8.05\times10^{-1}$~eV). All rovibrational energies of the diatomic H$_2$ are computed using the Fourier-Grid-Hamiltonian (FGH) method of Balint-Kurti and co-workers.\cite{MARSTON1989} The coupled Hamilton equations of motions in Jacobi coordinates are propagated using a Runge-Kutta 5 with adaptative time-step size propagator and finite difference formula of order three to compute numerically space derivatives.  The initial and final atom-distance was set to $25$ {\it a$_0$} and we constrain the relative error on the conservation of the total energy and total angular momentum below $10^{-6}$. We run batches of 40~000~trajectories for each energy point in order to reach an accuracy of less than $1\permil $\, for the lower energy points and $1\%$ for the upper energy limit.  Once trajectories are ended, the assignment of product quantum numbers ($v^{\prime}$,$j^{\prime}$) was done with the same procedure as described in ref. \citenum{DuncanMNRAS} which uses a histogram binning method.\cite{Truhlar1979}

The trajectory count in the histogram bin labeled $(v^{\prime},j^{\prime})$, denoted $N_r (v^{\prime},j^{\prime};E_c)$, allows us to determine the state-to-state cross-section for the reaction:
\begin{align}
	\sigma{\textrm{\scriptsize $v^{\prime},j^{\prime}\leftarrow v,j$}} (E_c)&=\pi b_{max}^2(E_c)\frac{N_r (v^{\prime},j^{\prime};E_c)}{N_{tot} (v,j;E_c)},
	\label{eq:xsec}
\end{align}
where $N_{tot} (v,j;E_c)$ is the total number of trajectories started in the ro-vibrational state $(v,j)$ and propagated with the collision energy $E_c$. The specific state rate coefficient was then obtained by summing state-to-state cross-sections overall final product ($v^{\prime}, j^{\prime}$) states:
\begin{align}
	\sigma{\textrm{\scriptsize $v,j$}} (E_c)&=\sum_{v^{\prime},j^{\prime}}\sigma_{\textrm{\scriptsize $v^{\prime},j^{\prime}\leftarrow v,j$}}(E_c),
	\label{eq:xsec}
\end{align}
and the associated rate coefficient is computed as:
\begin{align}
	k_{\textrm{\scriptsize $v,j$}} (T)=&\left(\frac{8\kB T}{\pi\mu}\right)^{1/2}\frac{1}{(\kB T)^2} \nonumber \\
		&\times\int_{0}^{\infty}\sigma_{\textrm{\scriptsize $v,j$}}(E_c)E_ce^{-\frac{E_c}{\kB T}}dE_c
	\label{eq:rateconst}
\end{align}
where $\kB$ is Boltzmann's constant, $\mu$ is the reduced mass for the D$^+$--H$_2$ motion, and where the integration was carried out numerically. We recall that in our case since we are looking for low temperature the calculation is only done for the state ($v=0,j=0$).\\
The highest internal energy levels $(v^{\prime},j^{\prime})$ of HD product considered in our collisional energy range are: (0,20), (1,18), (2,15), (3,12), (4,8) since all others channels are closed.

\subsection{Ring polymer molecular dynamics method}
RPMD method is based on the imaginary-time path integral formalism and takes advantage of the isomorphism between the quantum statistical mechanics of the system and the classical statistical mechanics of a fictitious ring polymer.\cite{Manolopoulos_JCP_2004} The ring polymer is composed of several copies of the original system (beads) connected by harmonic springs.\cite{Chandler_JCP_1981} The method is essentially a classical molecular dynamical method in an extended ring polymer phase space. Although RPMD theory was derived in an {\em{ad hoc}} manner,\cite{Manolopoulos_JCP_2004} recently its connection to the exact quantum Kubo-transformed\cite{Kubo_JPSJ_1957} time-correlation function has been established via a Boltzmann-conserving ``Matsubara dynamics'' (with explicit terms that are discarded).\cite{Hele_JCP_2015, Hele_JCP_2015_1} Matsubara dynamics considers the evolution of the low frequency, smooth ``Matsubara'' modes of the path integral,\cite{Matsubara_PTP_1955} but is computationally too expensive to be applied on complex systems. One can also easily recognize the connection between RMPD and the classical and quantum transition state theories (TST's) and the instanton theory under certain circumstances.\cite{Manolopoulos_Annu_Rev_Phys_Chem_2013, Hele_JCP_2013}

RPMD theory provides reliable estimates for the correlation functions responsible for thermal rate coefficients as demonstrated when compared with the experiment and quantum mechanical (QM) calculations,\cite{Suleimanov_JPCA_rev_2016} illustrated by its high level of accuracy. One of the most important features of RPMD reaction rate is in the fact that it is independent of the choice of the transition state diving surface,\cite{Manolopoulos_JCP_2005_2} which is beneficial for complex multidimensional reactions, where it is often difficult to locate the exact transition state. This aspect makes RPMD method considerably different from various TST methods wherein it leads to the unwanted parameter tuning due to its semiclassical nature.\cite{Manolopoulos_Annu_Rev_Phys_Chem_2013} The RPMD theory, in practice, can calculate the thermal rate coefficients of reactions involving hundreds of atoms.\cite{Miller_PNAS_2011} 

The RPMD reaction rate theory\cite{Manolopoulos_JCP_2005, Manolopoulos_JCP_2005_2} can yield exact QM result under certain conditions such as when the bath temperature is high (the ring polymer turn into a single bead), for short-time evaluation (upper bound to RPMD rate), for potential energy imitating a harmonic potential.\cite{Manolopoulos_Annu_Rev_Phys_Chem_2013} Owing to its simplicity and efficiency, RPMD theory has been applied to many bimolecular reactions such as: H$_2$ + H and their isotopologues \cite{Suleimanov_JCP_2009, Suleimanov_JPCL_2012, Suleimanov_PCCP_2013, Suleimanov_JPCL_2014_1, Suleimanov_JCP_2015}/C\cite{Suleimanov_JCP_2014,Suleimanov_JPCL_2015}/N\cite{Suleimanov_JPCL_2014}/O\cite{Suleimanov_JPCL_2014,Suleimanov_JPCA_2017}/F\cite{Suleimanov_JCP_2009, Suleimanov_JCP_2015}/Cl\cite{Suleimanov_JPCA_2014_2}/S\cite{Suleimanov_JCP_2014}/OH,\cite{Suleimanov_PCCP_2017_2} C + CH$^+$\cite{Suleimanov_JPCA_2016}/D$_2$,\cite{Suleimanov_PCCP_2017_3} F + NH$_3$,\cite{Suleimanov_JPCA_2014}  Cl + HCl\cite{Suleimanov_JCP_2009}/O$_3$,\cite{Suleimanov_PCCP_2014} CH$_4$ + H and its isotopologues\cite{Suleimanov_JCP_2011, Suleimanov_JCP_2013, Suleimanov_JCP_2015} /$^4$He$\mu$\cite{Suleimanov_JPCB_2015}/ O\cite{Suleimanov_JPCL_2013, Suleimanov_JPCA_2014_1}/OH\cite{Suleimanov_JCP_2013_1, Suleimanov_JPCB_2016}/CN\cite{Suleimanov_PCCP_2017} among others. A very recent paper show its applicability in determining the rate coefficients of key reactions in the interstellar medium.\cite{Suleimanov_SA_2018} As proved through extensive comparison with QM calculations, the RPMD rate coefficients are also reliable at intermediate temperatures.\cite{Suleimanov_JPCA_rev_2016} Since the quantum Boltzmann operator is treated accurately in the RPMD theory, it is able to map the zero-point energy (ZPE) effects\cite{Suleimanov_JPCL_2012} precisely along the entire reaction pathway.\cite{Suleimanov_JCP_2014, Suleimanov_JPCL_2014, Suleimanov_JPCL_2018} It is also more accurate than other TST approximate methods in the deep quantum tunneling regime.\cite{Suleimanov_JPCL_2014_1}

Determination of the rate coefficients for the barrierless complex-forming reactions must require recognizing two stumbling blocks. First, the presence of small free energy barrier that is generally difficult to locate. These energy barriers often lie in the asymptotic region. Second, if the intermediate is sufficiently long-lived, then an active recrossing effect is anticipated. RPMD can circumvent this issue because in essence, the recrossing factor is formally included in the method and the reaction rate is independent of the choice of the diving surface. As a result, RPMD method shows impressive accuracy for triatomic insertion reactions such the reaction between H$_2$ and N,\cite{Suleimanov_JPCL_2014} O,\cite{Suleimanov_JPCL_2014, Suleimanov_JPCA_2017} S,\cite{Suleimanov_JCP_2014} and C.\cite{Suleimanov_JCP_2014, Suleimanov_JPCL_2015} For the latter, excellent agreement with experiment has been achieved at very low temperatures.\cite{Suleimanov_JPCL_2015} RPMD rate coefficients are in very good agreement with the ones obtained with the accurate quantum dynamical (QD) calculations\cite{Suleimanov_JPCA_rev_2016} as well as with the experiment. RPMD method also proved to be very successful for a proper description of the ion-molecule reactions, as evident when the theory was applied to the  C + CH$^+$ reaction for the first time.\cite{Suleimanov_JPCA_2016} The authors, however, did find a large discrepancy between the RPMD and QCT results, particularly at the low temperature (20 $-$ 100 K). This difference has been attributed to the fact the QCT method does not take account of the changes in ZPE along the reaction coordinate and the error amplifies for barrierless channels at low temperature having deep potential well, as true in the case for C + CH$^+$ reaction. The title reaction being a prototype for the simplest ion-molecule proton transfer reaction with a barrierless ground state ($\tilde{X}~^1A^{\prime}$) entrance channels\cite{Aguado_JCP_2000} presents another opportunity to test the accuracy of the RPMD method at such low temperature regime of interest and made a comparison with QCT simulations on the same PES. We particularly focus on the evolution of the rate coefficient with the temperature by comparison of our actual simulations with the results obtained recently using more sophisticated quantum method such as TIQM and SQM (on a more recent but most time consuming PES). 

Below we summarize the RPMDrate code and its underlying theory. For a more detailed description of the RPMD rate theory and the explanation on the input parameters, the authors encourage the reader to ref.~\citenum{Suleimanov_JCP_2011} and \citenum{Suleimanov_CPC_2013} and a recent review article by one of us.\cite{Suleimanov_JPCA_rev_2016} For a typical gas phase bimolecular reaction,
\begin{equation*}
A + B \to {\text{intermediate~state}} \to {\text{products}}
\end{equation*}\label{eq:hamiltonian}
between reactants $A$ and $B$ with masses $m_A$ and $m_B$ respectively, the ring polymer Hamiltonian can be written in atomic Cartesian coordinates (in atomic unit) as:
\begin{eqnarray}
H_n(\boldsymbol{p},\boldsymbol{q})&=&\sum_{i=1}^{N}\sum_{j=1}^{n}\left(\frac{|\boldsymbol{p}_i^{(j)}|^2}{2m_i}+\frac{1}{2}m_i\omega_n^2|\boldsymbol{q}_i^{(j)}-\boldsymbol{q}_i^{(j-1)}|^2\right) \nonumber\\ & &+ \sum_{j=1}^{n}V(\boldsymbol{q}_1^{(j)},\boldsymbol{q}_2^{(j)},...,\boldsymbol{q}_{N}^{(j)}).
\end{eqnarray}
In this ring polymer system of $N$ atoms, the $i^{\text{th}}$ quantum particle of mass $m_i$ is represented by a necklace with $n$ classical beads connected by a harmonic potential with force constant $\omega_n$ ($=\beta_n\hbar$). $\beta_n \equiv \beta/n$ is the reciprocal temperature of the system $\beta=1/\kB T$. $\boldsymbol{p}_i^{(j)}$ and $\boldsymbol{q}_i^{(j)}$ are the momentum and position vectors of the $j^{\text{th}}$ bead in the ring polymer necklace of atom $i$, respectively.

The RPMD rate coefficient at temperature $T$, $\kRPMD(T)$, can be written in terms of the Bennett-Chandler factorization\cite{Bennett_ACC_1978, Chandler_JCP_1978} scheme as:\cite{Suleimanov_CPC_2013}
\begin{equation}\label{eq:rpmdrate}
\kRPMD(T)=\kQTST(T;\xi^{\ddagger}) \kappa(t\to \infty;\xi^{\ddagger}).
\end{equation}
In the present study, the thermal rate coefficients for the title reaction have been calculated at four different temperatures: 20, 50, 75 and 100 K. We have used a larger number of beads for temperatures below 50 K  (160 compared to 128 for $T> 50$ K) for convergence and scaled it inversely to the temperature when increasing it. The first term of eqn.~(\ref{eq:rpmdrate}) contributes towards the static part of $\kRPMD$ and is referred to as the centroid-density quantum transition-state theory\cite{Gillan_PRL_1987, Gillan_JPC_1987, Voth_JCP_1989} (QTST) rate coefficient.\cite{Manolopoulos_JCP_2005_2} $\kQTST(T;\xi^{\ddagger})$ is evaluated at the transition state $\xi^{\ddagger}$ along the reaction coordinate $\xi(\bar{\boldsymbol{q}})$ (or for barrierless reactions it is advantageous to evaluate at the free energy maximum). $\xi(\bar{\boldsymbol{q}})$ can be considered as an interpolating function that relates two dividing surfaces, $s_0$ (in the asymptotic reactant valley) and $s_1$ (in the intermediate state region) by:
\begin{equation}\label{eq:rxncord}
\xi(\bar{\boldsymbol{q}})=\frac{s_0(\bar{\boldsymbol{q}})}{s_0(\bar{\boldsymbol{q}})-s_1(\bar{\boldsymbol{q}})},
\end{equation}
such that $\xi \to 0$ as $s_0 \to 0$  (reactant) and $\xi \to 1$ as $s_1 \to 0$ (intermediate). It is clear from eqn.~(\ref{eq:rxncord}) that $\kQTST(T;\xi^{\ddagger})$ depends on the position of the dividing surface and is determined by the static equilibrium properties. It can also be calculated from the centroid potential of mean force (PMF)\cite{Suleimanov_JCP_2009, Suleimanov_JCP_2011, Suleimanov_CPC_2013} along the reaction coordinate, $W(\xi)$ by the following equation:
\begin{equation}\label{eq:kqtst}
\kQTST(T;\xi^{\ddagger})=4\pi R_{\infty}^2\left(\frac{1}{2\pi\beta\mu_R}\right)^{1/2}e^{-\beta\left[W(\xi^{\ddagger})-W(0)\right]}.
\end{equation}
Here, $R_{\infty}$ is the asymptotic distance between the reactants $A$ and $B$ and is chosen large enough to neglect the interaction between the reactants. At low temperatures, the results can be very sensitive to this parameter and therefore we have analyzed several values of $R_{\infty}$. $\mu_R$ is the reduced mass of the reactants,
\begin{equation}\label{eq:redmass}
\mu_R=\frac{m_A + m_B}{m_A \times m_B}.
\end{equation}
The free energy difference $\left[W(\xi^{\ddagger})-W(0)\right]$ in eqn.~(\ref{eq:kqtst}) is calculated by the umbrella integration procedure\cite{Thiel_JCP_2005, Thiel_JCP_2006} along the reaction coordinate $\xi$. To calculate the PMF profiles, $\xi$ has been divided into 111 equally spaced windows within the range [$-$0.05, 1.05], with each window separated by a width of 0.01. As an intermediate state, we have selected the position of the complex well. However, since it is preferable to initiate the recrossing dynamics calculations at the top of the PMF profile, we have limited the range to [$-$0.05, 0.80 (75, 100 K)/0.70 (50 K)/0.30 (20 K)] when an approximate position of the maxima was located. For each window centered at $\xi_i$, 100 RPMD trajectory calculations have been performed for 40 ps with time-step of 0.1 fs under constraint by adding a harmonic potential of form $K(\xi-\xi_i)^2$ to the Hamiltonian of the system in the presence of an Andersen thermostat.\cite{Andersen_JCP_1980} Here, $K$ is the force constant, chosen (2.72 ($T$/K) eV) such that it is large enough to merely explore the surroundings of $\xi_i$ and simultaneously small enough to allow overlapping between neighboring $\xi$ distributions. The first 12 ps are used for thermalization for trajectory calculation.

The second term of eqn.~(\ref{eq:rpmdrate}), $\kappa(t\to \infty;\xi^{\ddagger})$, is the long-time limit of a time-dependent ring polymer transmission coefficient (the ring polymer recrossing factor). It is essentially a dynamical correction to the centroid density QTST rate coefficient $\kQTST(T;\xi^{\ddagger})$ and accounts for the recrossings at the top of the free-energy barrier ($\xi^{\ddagger}$) until a ``plateau'' time\cite{Chandler_JCP_1978} is reached. This factor ensures that the RPMD rate coefficient is independent of the choice of dividing surfaces, $s_0(\bar{\boldsymbol{q}})$ and $s_1(\bar{\boldsymbol{q}})$ and consequently $\xi(\bar{\boldsymbol{q}})$, by counterbalancing $\kQTST(T;\xi^{\ddagger})$. It is expressed as the ratio between two-flux side correlation functions:
\begin{equation}\label{eq:kappa}
\kappa(t \to \infty;\xi^{\ddagger})=\frac {c_{fs}^{(n)}(t \to \infty;\xi^{\ddagger})}{c_{fs}^{(n)}(t \to 0_+;\xi^{\ddagger})}.
\end{equation}
$\kappa(t\to \infty;\xi^{\ddagger})$ is calculated by running trajectories initiated at the maximum of the PMF that corresponds to the reaction coordinate $\xi^{\ddagger}$. For the title reaction, the PMF maximum shifts gradually toward smaller $\xi$ as the temperature of the system is lowered. As mentioned previously, we optimize $\xi^{\ddagger}$ for each temperature and terminate umbrella integration before entering the deep well in the PMF profile so as to minimize the recrossings (faster convergence) and time required to reach plateau value of $\kappa$. For a considerable sampling of initial conditions, a long ``parent'' trajectory of length 4 ns (20, 50 and 75 K) or 2 ns (100 K) has been carried out after a thermalization period of 20 ps in the presence of an Andersen thermostat,\cite{Andersen_JCP_1980} with its centroid pinned at $\xi^{\ddagger}$. Configurations of the parent trajectory are sampled successively for each 2 ps period that would serve as initial positions for the ``child" trajectories that are used to compute the recrossing factor. For each initial position, there are either 50 (20, 50 and 75 K) or 100 (100 K) separate child trajectories that are spawned with different initial momenta sampled from a Boltzmann distribution. These trajectories are then propagated for 3 ps without the thermostat or the dividing surface constraint to ensure that the transmission coefficients reach plateau values.

The RPMD calculations have been performed using the RPMDrate code.\cite{Suleimanov_CPC_2013} The input parameters for the code are summarized in Table~\ref{tab:parameters}. They are similar to the ones used in the previous RPMDrate study of similar ion-molecule reaction (C + CH$^+$)\cite{Suleimanov_JPCA_2016} as well as several previous studies of insertion chemical reactions.\cite{Suleimanov_JCP_2014, Suleimanov_JPCL_2014, Suleimanov_JPCL_2015}

\begin{table}[h]
\scriptsize
\caption{Input parameters for the RPMD calculations on the D$^+$ + H$_2$ reaction$^{\text\em {a}}$}
\label{tab:parameters}
\begin{center}
\begin{tabular}{lll} 
\hline
Parameter &{PES} & Explanation\\
\cline{2-3}
& $\tilde{X}~^1A^{\prime}$  & Ref.~\citenum{Aguado_JCP_2000} \\
\hline
\multicolumn{3}{l}{Command line parameters} \\ 
\hline
\ttfamily{Temp}         & {100}      &       Temperature (K)  \\
                                & {75}     &         \\
                                &{50}       &         \\
                                 & {25}       &         \\
\ttfamily{Nbeads}       &{128 (75, 100 K)} & Number of beads \\
&	160 (20, 50 K) &\\
\hline
\multicolumn{3}{l}{Dividing surface parameters} \\ 
\hline
$R_\infty $  & 12 \AA\,(100, 75, 50 and 20 K) &  Dividing surface parameter (distance) \\
& 20 \AA\,(50 and 20 K) & \\
& 30 \AA\,(50 and 20 K) & \\
$N_{\rm bonds}$       &  2    & Number of forming and breaking bonds \\
$N_{\rm channel}$     &  2   & Number of equivalent product channels \\
H & (0.78, $-$1.22, 0.50) & Cartesian coordinates (x, y, z) of the \\
H &(1.60, $-$1.16, 0.51) & intermediate geometry (\AA) \\ 
D & (1.12, $-$0.66, $-$0.03) & \\ 

\hline
\ttfamily{Thermostat}  & `Andersen'   &  Thermostat option \\
\hline
\multicolumn{3}{l}{Biased sampling parameters} \\ 
\hline
$N_{\rm windows}$  & 111        & Number of windows \\
$\xi _1$            & $-$0.05    & Center of the first window \\
$d\xi $              &  0.01      & Window spacing step      \\ 
$\xi _N$             &  1.05    & Center of the last window \\
$dt$  &  0.0001   &  Time step (ps) \\
$k_i$  &  2.72  & Umbrella force constant ((T/K) eV) \\
$N_{\rm trajectory}$ & 100  & Number of trajectories \\ 
$t_{\rm equilibration}$ &  12    & Equilibration period (ps)  \\
$t_{\rm sampling}$ & 40    & Sampling period in each trajectory (ps) \\
$N_i$   & $4\times 10^7$  &  Total number of sampling points \\
\hline
\multicolumn{3}{l}{Potential of mean force calculation} \\ 
\hline
$\xi _0$            & $-$0.020   & Start of umbrella integration  \\
$\xi ^{\ddagger}$             & 0.80 (75, 100 K)$^{\text\em {b}}$   & End of umbrella integration \\
                                       &  0.70 (50 K)$^{\text\em {b}}$ &    \\
                                       &  0.30 (20 K)$^{\text\em {b}}$ &    \\
$N_{\rm bins}$             & 4999   &  Number of bins \\
\hline
\multicolumn{3}{l}{Recrossing factor calculation} \\ 
\hline
$dt$  & 0.0001  &Time step (ps) \\
$t_{\rm equilibration}$  & 20   & Equilibration period (ps) in the constrained \\
&&(parent) trajectory\\
$N_{\rm totalchild}$  & 100000  & Total number of unconstrained (child)\\
&& trajectories \\
$t_{\rm childsampling}$  & 2  & Sampling increment along the parent\\
&& trajectory  (ps) \\
$N_{\rm child}$  & 100 (100 K) &  Number of child trajectories per one\\
& 50 (20, 50, 75 K) & initially constrained configuration\\
$t_{\rm child}$  & 3   &Length of child trajectories (ps) \\
\hline
\end{tabular}\\
\end{center}
$^{\text\em {a}}$ {The explanation of the format of the input file can be found in the RPMDrate code manual (\href{http://rpmdrate.cyi.ac.cy}{http://rpmdrate.cyi.ac.cy}).} $^{\text\em {b}}$ {Detected automatically by RPMDrate.} 
\end{table}

The present RPMD calculations are also based on the adiabatic full-dimensional global PES for the ground $\tilde{X}~^1A^{\prime}$ state of the H$_3^+$ system, developed by Aguado and co-workers (ARTSP PES). 

\section{Results and discussion}\label{sec:result}
\begin{figure}[h]
   \centering
   \includegraphics[width=0.47\textwidth]{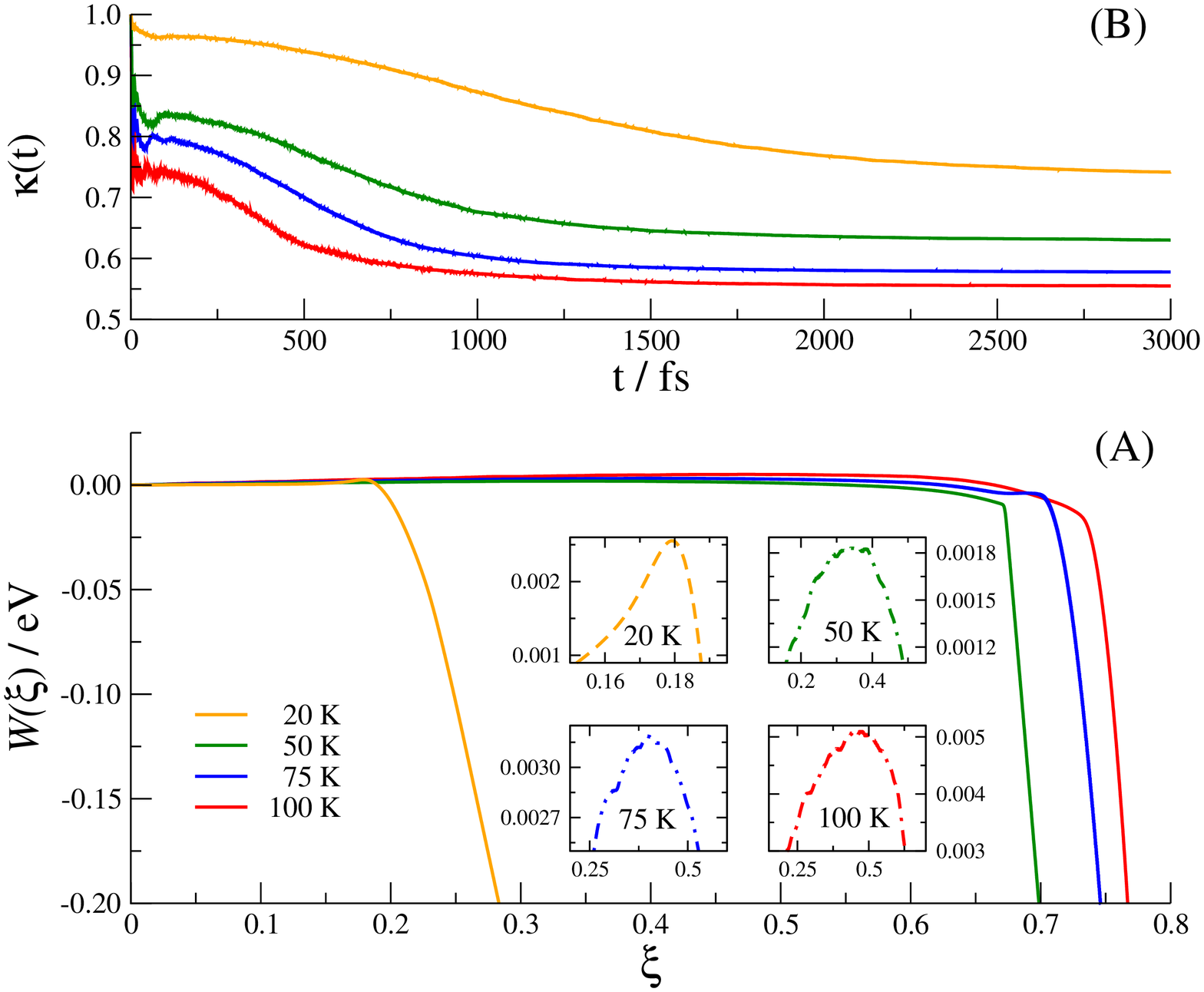} 
   \caption{(A) Variation of RPMD potential of mean force, $W(\xi)$, (in eV) along the reaction coordinate $\xi$ for the temperature range 20 $-$ 100 K. Inset plots are magnification of the barrier region in the PMF profile. (B) RPMD time dependent transmission coefficient, $\kappa(t)$, in the temperature range 20 $-$ 100 K. The legends correspond to both (A) and (B). The asymptotic distance between the reactants, $R_\infty $, in all calculations, is set at 12 \AA.}
   \label{fig:PMF_RF}
\end{figure}

At first, we will summarize the results obtained with $R_\infty $ = 12 \AA, which is a common parameter in the RPMD simulations at all temperatures. The variation of RPMD potential mean force $W(\xi)$ along the reaction coordinate $\xi$ at $T$ = 20, 50, 75 and 100 K is plotted in Fig.~\ref{fig:PMF_RF}(A). The PMF profiles for all temperature range have similar characteristics and are essentially barrierless. However, a closer inspection on each of these PMF profiles shows the existence of a very low thermodynamic barrier (see inset plots of  Fig.~\ref{fig:PMF_RF}(A)) that gradually shifts towards smaller $\xi$ as the temperature of the system is lowered. At $T$ = 20 K, the maximum in the PMF profile is situated around $\xi$ = 0.18, thus emphasizing the importance of long-range interaction as the temperature decreases. The maximum barrier height (5.09$\times 10^{-3}$ eV) is obtained for $T$ = 100 K at $\xi$ = 0.47. It is interesting to note that the barrier height becomes smaller with low $T$, with one exception being at $T$ = 20 K. At 20 K, the barrier height (2.55$\times 10^{-3}$ eV) is marginally greater than at 50 K (1.84$\times 10^{-3}$ eV). The difference can be due to the convergence of the umbrella integration procedure. In these PMF profiles, a deep well (fast decrease in the PMF), with a well-depth of at least 0.2 eV, is also present, an indication that the title reaction proceeds through a complex formation.\cite{Suleimanov_JCP_2014} The reaction coordinate where the formation of the deep well takes place moves towards the higher value of $\xi$ with the rising temperature, thus approaching the classical limit of the underlying PES. At 20 K, the well is initiated around $\xi \approx 0.19$, which is very close to the potential barrier ($\xi \approx 0.18$). However, for the remaining temperatures, the well starts to form around $\xi \approx 0.67-0.74$, which is at least 0.27 reaction coordinate further away from the barrier. The steepness of these wells also increases with temperature. For instance, the free energy gradient around the well at 20 K is found to be $\approx -$2.41 eV, while that for 100 K is $\approx -$9.94 eV. These PMFs do however follow a similar barrierless characteristic for other insertion type bimolecular reactions of H$_2$ molecule, such as with C($^1D$) and S($^1D$).\cite{Suleimanov_JCP_2014} As pointed out previously, the RPMD trajectories were not propagated further from the deep potential well.

The time-dependent transmission coefficients, $\kappa(t)$, for all the temperatures considered in our study are depicted in Fig.~\ref{fig:PMF_RF}(B). The corresponding plateau value of transmission coefficient, $\kappa(t\to\infty)$, is provided in Table~\ref{tab:results}. It clear that recrossings do take place at all temperatures and therefore it plays an important role by significantly altering the final value of the RPMD rate coefficient, as compared to the QTST rate. After a long propagation (2 $-$ 3 ps) and initial oscillations (up to 40 fs), $\kappa(t)$ reaches a constant value. The oscillations are due to the choice of the dividing surface and may correspond to the H$_2$ stretching vibration\cite{Suleimanov_JCP_2009} on the ARTSP PES. Very long propagation time is required to converge the transmission coefficients as expected, since the title reaction proceeds through a complex-forming reaction mechanism.\cite{Suleimanov_JPCL_2014} $\kappa(t)$ at 100 K reaches the plateau in the least amount of time than for any other temperatures. This is apparently due to its greater activity in entering and re-entering the complex forming zone, aided by higher temperature.\cite{Suleimanov_JCP_2014} For all cases, $\kappa(t\to\infty)$ lie within the range of 0.74 $-$ 0.55, which is substantially below 1, and consequently justifies the need of including this real-time recrossing correction factor in the final rate coefficient. From the values of $\kappa(t\to\infty)$, it is very evident that the recrossing tends to increase with increasing temperature. For {\textit{e.~g.}}, $\kappa$ is 0.55  at 100 K, while at 20 K, it increases to 0.74. The behavior of the recrossing factor with temperature $T$ is similar to the ones observed previously for the reactions involving H$_2$ molecule with C\cite{Suleimanov_JPCL_2015} and O.\cite{Suleimanov_JPCA_2017} We know that the temperature dependence of $\kappa(t)$ is related to the observed PMF profile. For instance, in the O + H$_2$ reaction, we observe a tiny barrier in the PMF profile calculated over the $1~^1A^{\prime}$ surface.\cite{Suleimanov_JPCA_2017} As a result, the temperature dependence of the recrossing factor was the same, it increases from 0.30 at 300 K to 0.54 at 50 K. In contrast, the PMF profile of the first excited state ($1~^1A^{\prime\prime}$ state) has more sizable free energy barrier (up to $\sim$0.16 eV) along the reaction coordinate. The recrossing factor of this state now has opposite temperature dependence, it decreases from 0.33 at 300 K to 0.01 at 50 K. Similar explanation can be given to the seemingly opposite behavior of $\kappa$ with $T$ for the $^1A^{\prime}$ and $^1A^{\prime\prime}$ states of the C + H$_2$ insertion reaction.\cite{Suleimanov_JPCL_2015}

\begin{table}[!htbp]
\caption{Summary of the rate calculations for the D$^+$ + H$_2$ reaction at temperatures ($T$) 20, 50, 75, and 100 K: $\kQTST$ $-$ centroid-density quantum transition state theory rate coefficient; $\kappap$ $-$ ring polymer transmission coefficient; $\kRPMD$ $-$ ring polymer molecular dynamics rate coefficient; $\kQCT$ $-$ quasi-classical trajectory method rate coefficient; $\kTIQM$ $-$ time-independent quantum method rate coefficient; $\kSQM$ $-$ statistical quantum mechanical rate coefficient; $\kST$ $-$ statistical theory rate coefficient. The values of the rate coefficients are reported in 10$^{-9}$ cm$^3$ s$^{-1}$. The asymptotic distance between the reactants, $R_\infty $, in all RPMD calculations, is set at 12 \AA}
\label{tab:results}
\begin{center}
\setlength{\tabcolsep}{.5em}
\begin{tabular}{llllllll}
\hline
$T ({\text{K}})$& $\kQTST$ & $\kappap$ & $\kRPMD$$^{\text\em {a}}$ & $\kQCT$$^{\text\em {a}}$ & $\kTIQM$$^{\text\em {b}}$ & $\kSQM$$^{\text\em {b}}$ & $\kST$$^{\text\em {c}}$ \\
\hline
20 & 1.81 & 0.74 & 1.34 & 1.38 &1.48 & 1.79 & $-$\\
50 & 3.02 & 0.63 & 1.91 & 1.53  &1.59 & 1.83 & 1.48\\
75 & 3.47 & 0.58 & 2.00 & 1.57 &1.63 & 1.85 &1.54\\
100 & 3.63 & 0.55 & 2.01 & 1.60 &1.66 & 1.86 & 1.57\\
\hline
\end{tabular}
\end{center}
$^{\text\em {a}}$ {Present study.} $^{\text\em {b}}$ {Ref.~\citenum{Scribano_JPCA_2014}.} $^{\text\em {c}}$ {Ref.~\citenum{Gerlich_PSS_2002}.}
\end{table}

Table~\ref{tab:results} compares the QTST, RPMD and QCT rate coefficients for $T$ = 20 $-$ 100 K with the previous theoretical calculations.\cite{Gerlich_PSS_2002, Scribano_JCP_2013, Scribano_JPCA_2013} It is quite evident that both $\kRPMD$ and $\kQCT$ marginally increases as $T$ increases. However, the change in the value of $\kQTST$ is much more rapid, particularly when the temperature increases from 20 K to 50 K. The calculated difference in the value of the RPMD rate coefficients within the entire temperature regime is up to 0.67$\times$10$^{-9}$ cm$^3$ s$^{-1}$, while that for the QTST rate coefficient is 1.82$\times$10$^{-9}$ cm$^3$ s$^{-1}$. On the other hand, this difference is much smaller in $\kQCT$ values (0.22$\times$10$^{-9}$ cm$^3$ s$^{-1}$). Therefore, $\kRPMD$ has although weak but correct temperature dependence, as noted by previous studies with TIQM and SQM methods.\cite{Scribano_JCP_2013, Scribano_JPCA_2013,Scribano_JPCA_2014} This is also consistent with other barrierless reactions, such as the reaction between H$_2$ and C/O/S.\cite{Suleimanov_JCP_2014,Suleimanov_JPCA_2017} Though the rate coefficients for these reactions are less sensitive to temperature (within $\approx 10^{-11}-10^{-12}$ cm$^3$ s$^{-1}$) than the present reaction. At 50 K and above, the rate coefficient does not change significantly. At 20 K, $\kRPMD$ is comparatively much smaller than other temperatures. This is a direct consequence of smaller value of the QTST rate coefficient rather than due to large recrossings, as discussed previously. 

From Table~\ref{tab:results}, it evident that the $\kRPMD$ is always greater than those obtained by QCT ($\kQCT$), ST ($\kST$), TIQM ($\kTIQM$) and SQM ($\kSQM$) methods at all temperatures except at 20 K. They all do however follow the same temperature trend, {\textit{i.~e.}}, they marginally increase with temperature. However, both $\kSQM$ and $\kST$ have the least dependence on $T$, as it changes only an amount of $\sim$0.07 $-$ 0.09$\times$10$^{-9}$ cm$^3$ s$^{-1}$ within the entire temperature regime. Although the maximum change in the $\kTIQM$ value (0.18$\times$10$^{-9}$ cm$^3$ s$^{-1}$) within the temperature range of this present study is marginally greater than for both $\kSQM$ and $\kST$, it is much smaller than for $\kRPMD$. It is interesting to note that the rate coefficients in each dynamical method start to form a plateau around 50 K. The RPMD results fall within 4 $-$ 25\% deviation with those obtained with QD for a particular temperature. RPMD deviation from the SQM calculations (4 $-$ 25\%) is a little greater than from the TIQM ones (9 $-$ 23\%). The maximum difference (0.45$\times$10$^{-9}$ cm$^3$ s$^{-1}$) between the RPMD and QD rate coefficient is obtained at $T$ = 20 K.  A part of this discrepancy between RPMD and TIQM/SQM results may arise from the difference between the potential energy surfaces used in these studies (ARTSP\cite{Aguado_JCP_2000} and VLABP\cite{Velilla_JCP_2008} PES) correspondingly. The long-range interaction potential has been more accurately defined in the VLABP PES. Moreover, it has been reported that RPMD may slightly overestimate the rates of barrierless reactions.\cite{Suleimanov_JCP_2014,Suleimanov_JPCL_2014}

It has been previously reported that the rate of D$^+$ + H$_2$ reaction depends to a larger extent on the asymptotic distance on the entrance channel of the reactants.\cite{Scribano_JPCA_2013, Scribano_JPCA_2014} As noted earlier, the modified version of ARTSP PES (VLABP PES) can correctly describe the long-range electrostatic potential by including an analytical representation of this interaction. It has been found that if the reactants are not largely separated at the asymptote, the corresponding dynamical calculation may incur errors in the final value of the rate coefficients despite using more accurate VLABP PES.\cite{Scribano_JPCA_2013} This is particularly true at low temperatures since the long-range interaction potential becomes more important for the QD calculations at low collision energies. For example: when the asymptotic Jacobi distance in  SQM and TIQM calculations is increased from 15 $a_0$ to 70 $a_0$ and 30 $a_0$ to 40 $a_0$ respectively, the thermal rate coefficient increases substantially within the temperature range of 10 $-$ 100 K.\cite{Scribano_JPCA_2013,Scribano_JCP_2013} This increase is more pronounced at lower temperature range, for {\textit{e.~g.}},  27\% $-$ 145\% increase for $T$ $\le$ 50 K compared to 15\% $-$ 21\% increase for $T$ $>$75 K. As a consequence, the rate coefficient emerges to be less dependent on the temperature than previously calculated and remain almost constant throughout the studied temperature range.\cite{Scribano_JPCA_2014} 

\begin{figure}[h]
   \centering
   \includegraphics[width=0.45\textwidth]{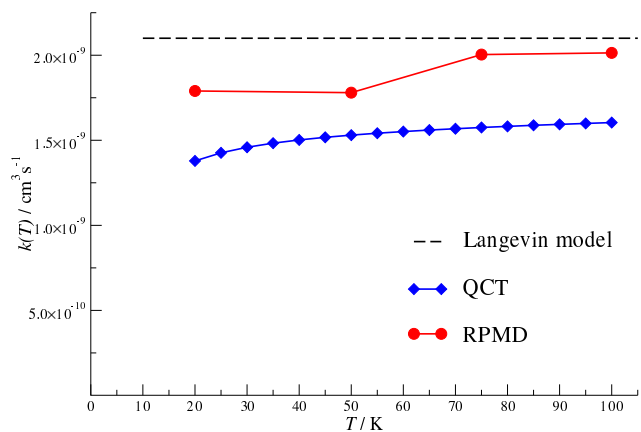} 
   \caption{Rate coefficients $k(T)$ (in cm$^3$ s$^{-1}$) for the D$^+$ + H$_2$ reaction as a function of temperature, $T$ (in K) calculated using ARTSP PES of Aguado {\textit{et al.}}\cite{Aguado_JCP_2000} Dashed black line: Langevin rate coefficient; solid blue line with diamonds: quasi-classical trajectory (QCT) rate coefficient; solid red line with circles:  ring polymer molecular dynamics (RPMD) rate coefficient with $R_{\infty}$ = 12 \AA\, (75 K and 100 K) / 20 \AA\, (20 K and 50 K). $R_\infty $ is the  asymptotic distance between the reactants.}
\label{fig:rate}
\end{figure}

To examine the presence of a similar dependence of $\kRPMD$ on the asymptotic distance of the reactants, we have performed two additional RPMD simulations with a longer $R_{\infty}$ (20 \AA \, and 30 \AA) at both 20 K and 50 K. We have kept the same parameters as described in Table~\ref{tab:parameters}, except the force constant $K$ has been increased from 2.7 (T/K) eV to 6.8 $-$ 20.4 (T/K) eV to sample a considerable number of distributions at long separations. Moreover, in each of these calculations, the child trajectories have been propagated for a longer period ($t_{\rm child}$ = 8 ps). The new PMF profiles have similar characteristics to those obtained earlier as in Fig.~\ref{fig:PMF_RF}(A). However, for $R_{\infty}$ = 30 \AA,\, the thermodynamic barrier height can slightly increase up to a maximum of 1.8$\times 10^{-2}$ eV. The comparison between the plateau value of the transmission coefficients, $\kappap$, with varying $R_{\infty}$ at $T$ = 20 and 50 K are  tabulated in Table~\ref{tab:rinfinity_results}. It is clear that the recrossings tend to increase with the increasing $R_{\infty}$, indicating that the complex decays back to the reactant channel more comfortably.  The corresponding rate coefficients are  compared in Table~\ref{tab:rinfinity_results} along with the $\kQTST$ values. It is evident that $\kQTST$ increases with increasing $R_{\infty}$. The increase in $\kQTST$ is more obvious in all cases except at $R_{\infty}$ = 30 \AA\, and $T$ = 20 K. The most striking feature in the new values of $\kRPMD$ is that it increases substantially at $T$ = 20 K. The percent increase in $\kRPMD$ value compared to that obtained at $R_{\infty}$ = 12 \AA\, is around 34 $-$ 40\%. At $T$ = 50 K, the changes in the rate coefficient are comparatively smaller. However, they do show opposite behavior to those obtained at $T$ = 20 K, {\textit{i.~e.}}, decrease with increasing $R_{\infty}$. This may be related to the fact that $\kappap$ decreases more rapidly than $\kQTST$ increases at 50 K compared to those at 20 K. The maximum difference between RPMD and SQM methods that was obtained at 20 K has been now virtually nullified. The primary observation of this inspection is that the rate coefficient is now less dependent on temperature (see  Fig.~\ref{fig:rate}), a remark corroborating earlier study by Gonz{\'a}lez-Lezana {\textit{et al.}}\cite{Scribano_JPCA_2014} Therefore, it is very important to choose the proper asymptotic distance between the reactants for the dynamical studies at low temperatures particularly when the collisional energy is small, as minor changes in this parameter can alter the rate coefficient quite significantly.    

\begin{table}[h]
\caption{Summary of the rate calculations for the D$^+$ + H$_2$ reaction calculated by using ARTSP PES of Aguado {\textit{et al.}}\cite{Aguado_JCP_2000} at temperatures ($T$) 20 and 50 K: $R_\infty $ $-$ asymptotic distance between the reactants in \AA; $\kQTST$ $-$ centroid-density quantum transition state theory rate coefficient; $\kappap$ $-$ ring polymer transmission coefficient; $\kRPMD$ $-$ ring polymer molecular dynamics rate coefficient. The values of the rate coefficients are reported in 10$^{-9}$ cm$^3$ s$^{-1}$}
\label{tab:rinfinity_results}
\begin{center}
\begin{tabular}{lclll}
\hline
$T ({\text{K}})$& $R_\infty $ (\AA)& $\kQTST$ & $\kappap$ & $\kRPMD$\\
\hline
\multirow{3}{*}{20} & 12 & 1.81 & 0.74 & 1.34\\
& 20 & 3.67 & 0.49 & 1.79\\
& 30 & 3.72 & 0.50 & 1.88\\
\cline{2-5}
\multirow{3}{*}{50} & 12 & 3.02 & 0.63 & 1.91\\
& 20 & 3.86 & 0.46 & 1.78\\
& 30 & 5.55 & 0.32 & 1.76\\
\hline
\end{tabular}
\end{center}
\end{table}

Both $\kRPMD$ and $\kQCT$ always have a smaller value than the rate coefficient calculated by the Langevin model (2.1$\times$10$^{-9}$ cm$^3$ s$^{-1}$). On the other hand, $\kQTST$ is always greater than the Langevin model, except at 20 K. Langevin model imposes several restrictions, such as: it implicitly assumes that the reactants are all in their ground rotational states.\cite{McCarroll_PS_2011} In practice, Langevin value is never reached since the collision complex also decays back to the reactant channel.\cite{Gerlich_PSS_2002} McCarroll modified the Langevin approach by introducing statistical mixing model using an isotropic long-range potential.\cite{McCarroll_PS_2011} Imposing the nuclear symmetry constraints, McCarroll found that the rate coefficients of both normal and para-H$_2$ converge to $\sim$2.1$\times$10$^{-9}$ cm$^3$ s$^{-1}$ at 10 K, which marginally increases with the temperature till 100 K. Rate coefficient of ortho-H$_2$ is rather small ($\sim$0.6$\times$10$^{-9}$ cm$^3$ s$^{-1}$) at 10 K and can be considered to be independent of temperature. 

Since there is a limited number of experimental measurements of the rate coefficients within the low temperature regime, we make a simple comparison with the FA measurements of Fehsenfeld {\textit{et al.}}\cite{Fehsenfeld_JCP_1974} At 80 K, the FA measurement reports the value of rate coefficient larger than 0.7$\times$10$^{-9}$ cm$^3$ s$^{-1}$. Although the precise FA value is rather uncertain, one may assume that it is much smaller than both $\kRPMD$ and $\kQCT$ values at 75 K. The FA rate coefficients within the temperature range of 200 $-$ 278 K remain constant at 1.0$\times$10$^{-9}$ cm$^3$ s$^{-1}$ with error bars of +0.5/$-$0.25$\times$10$^{-9}$ cm$^3$ s$^{-1}$. The FA measurements are apparently also in contradiction to those obtained with SIFT\cite{Henchman_JCP_1981} apparatus. For {\textit{e.~g.}}, the rate coefficients obtained by SIFT analysis cited to fall within (1.7 $-$ 2.2)$\times$10$^{-9}$ cm$^3$ s$^{-1}$ for the temperature range of 205 K to 295 K. Therefore, we argue that new experimental measurements are required to track the range in which rate coefficients are to be expected. This will then elucidate whether the present or previous theoretical estimates of the rate coefficients as well as the PES employed in these studies are valid and correct. 

\section{Conclusions}\label{sec:conclusion}
In this work, we have calculated the thermal rate coefficients of the D$^+$ + H$_2$ reaction within the temperature range of 20 $-$ 100 K by the ring polymer molecular dynamics (RPMD) and quasi-classical trajectory (QCT) methods. The calculations have been performed on the ground state ($\tilde{X}~^1A^{\prime}$) of the system, using the analytical potential energy surface developed by Aguado {\textit{et al.}}\cite{Aguado_JCP_2000}  Since the entrance channel of this reactions is virtually barrierless and the reaction evolves through the formation of a long-lived intermediate complex,\cite{McCarroll_PS_2011} it poses an interesting dynamical problem as such an active recrossing in dynamical simulations is expected. The RPMD method, which automatically includes the recrossing effects, shows the importance of recognizing this factor in the rate coefficient evaluation as it notably changes its final value. Both RPMD and QCT rate coefficients are in good agreement with earlier the TIQM and SQM studies.\cite{Scribano_JPCA_2014}  They fall within the range of  1.34 $-$ 2.01$\times$10$^{-9}$ cm$^3$ s$^{-1}$ and exhibit very low temperature dependence slightly decreasing when the temperature is decreased. 

In addition, we find that the parameter which is responsible for the asymptotic distance between the reactants ($R_{\infty}$) in the RPMD computational procedure has a huge influence on the resulting value of the rate coefficient, particularly at very low temperatures (20 $-$ 50 K). For instance, at 20 K, there is a 40\% increase in the rate coefficient value when the asymptotic distance has been increased from 12 \AA\, to 30 \AA. As a result, the reaction rate becomes almost invariant under the change in temperature within the studied regime. Therefore, this parameter must be correctly defined for the dynamical studies of the ion-molecule reactions at the low temperature, as they tend to have a deep potential well. This is due to the significant contribution of the long-range interaction part of the underlying PES at the low temperatures, which is not frequently encountered in the ambient-temperature regime of chemical reactivity. 

In conclusion, in the present study, we have corroborated the efficient and rigorous nature of the RPMD method for determining low temperature thermal rate coefficients for chemical reactions of astrochemical importance. We also hope that this work will stimulate future experimental measurements of the rate coefficients for the title reaction at low temperatures  due to the fact that there is a lack of experimental data below 100 K. 

\section*{Conflicts of interest}
There are no conflicts to declare.

\section*{Acknowledgements}
Y.V.S. and S.B. thanks the European Regional Development Fund and the Republic of Cyprus for support through the Research Promotion Foundation (Project Cy-Tera NEA ${\rm Y\Pi O\Delta OMH}$ / ${\rm \Sigma TPATH}$/0308/31) (Cy-Tera Project number: pro17b106s2). S.B. also acknowledges the support of the COST Action CM1401/STSM:40478 (Our Astro-Chemical History). The calculations have been performed at the CyI (Cy-Tera) supercomputer center and also with the support of the High Performance Computing Platform MESO@LR, financed by the Occitanie / Pyr\'en\'ees-M\'editerran\'ee Region, Montpellier Mediterranean Metropole and the University of Montpellier.



\balance

\renewcommand\refname{References}

\bibliography{bibliography} 
\bibliographystyle{rsc} 

\end{document}